\documentclass[times,doublespace]{wcmauth}

\usepackage{cite}
\usepackage{moreverb}
\usepackage[printonlyused]{acronym}
\graphicspath{{./}}
\usepackage[tight,footnotesize]{subfigure}
\newcommand\BibTeX{{\rmfamily B\kern-.05em \textsc{i\kern-.025em b}\kern-.08em
T\kern-.1667em\lower.7ex\hbox{E}\kern-.125emX}}

\def\volumeyear{2014}
\acrodef{PHY}{Physical layer}
\acrodef{BS}{Base Station}
\acrodef{CQI}{Channel Quality Information}
\acrodef{CoMP}{Coordinated Multi-Point}
\acrodef{LP}{Linear Program}
\acrodef{LIRA}{Long-term Inter-cell Resource Allocation}
\acrodef{LLS}{Long-term Lookback Scheduling}
\acrodef{LL}{Long-term Lookback}
\acrodef{LLPF}{Long-term Lookback Proportional Fair}
\acrodef{HLS}{HTTP Live Streaming}
\acrodef{LL-EXP}{Long-term Lookback Exponential}
\acrodef{LL-PF}{Long-term Lookback Proportional Fair}
\acrodef{EXP-F}{Exponential with Freezing}
\acrodef{EXP}{Exponential}
\acrodef{QoE}{Quality of Experience}
\acrodef{QoS}{Quality of Service}
\acrodef{MA}{Moving Average}
\acrodef{MR}{Maximum Rate}
\acrodef{RWP}{Random Way Point}
\acrodef{PF}{Proportional Fair}
\acrodef{RA}{Resource Allocation}
\acrodef{PFS}{Proportional Fair Scheduling}
\acrodef{MPFS}{Multi-cell Proportional Fair Scheduling}
\acrodef{IPF}{Inter-cell Proportional Fair}
\acrodef{MPF}{Multi-cell Proportional Fair}
\acrodef{MSE}{Mean Square Error}
\acrodef{SPF}{Single-cell Proportional Fair}
\acrodef{SPFS}{Single-cell Proportional Fair Scheduling}
\acrodef{QoS}{Quality of Service}
\acrodef{TCP}{Transmission Control Protocol}
\acrodef{TTI}{Time Transmission Interval}
\acrodef{UE}{User Equipment}
\acrodef{TDMA}{Time Division Multiple Access}
\acrodef{MAC}{Medium Access Control}
    \acrodef{LTE}{Long Term Evolution}
    \acrodef{HARQ}{Hybrid Automatic Repeat Request}
    \acrodef{STC}{Space-Time Coding}
    \acrodef{QAM}{Quadrature Amplitude Modulation}

\newcommand{\fref}[1]{Figure~\ref{#1}}
\newcommand{\sref}[1]{Section~\ref{#1}}

\newcommand{\eref}[1]{(\ref{#1})}
\newcommand{\tn}{\textnormal}

\begin{document}
\bibliographystyle{wileyj}
\runningheads{H.~Abou-zeid \textit{et al}.}{Lookback Scheduling for Long-Term QoS}
% \journalabb\ class file

\articletype{RESEARCH ARTICLE}
\title{A Lookback Scheduling Framework for Long-Term Quality-of-Service Over Multiple Cells}
%\footnotemark[2]
\author{
H.~Abou-zeid\affilnum{1}\corrauth, H.~S.~Hassanein\affilnum{1}, S. Valentin\affilnum{2}, and M.~F.~Feteiha\affilnum{1}
}
\address{\affil{1}{Queen's University, Kingston, ON, Canada, K7L 3N6}\\\affil{2}{Bell Labs, Alcatel-Lucent, Stuttgart, Germany}}
\corraddr{Hatem Abou-zeid, Telecommunications Research Lab, Electrical and Computer Engineering Department, Queen's University, Kingston, ON, Canada, K7L 3N6.\\E-mail: h.abouzeid@queensu.ca}

%\thanks{This work was made possible by a \emph{National Priorities Research Program} (NPRP) grant from the Qatar National Research Fund (Member of Qatar Foundation).}

\begin{abstract}
In current cellular networks, schedulers allocate wireless channel resources to users based on instantaneous channel gains and short-term moving averages of user rates and queue lengths. By using only such short-term information, schedulers ignore the users' service history in previous cells and, thus, cannot guarantee long-term \ac{QoS} when users traverse multiple cells with varying load and capacity. In this paper, we propose a new \ac{LLS} framework, which extends conventional short-term scheduling with long-term \ac{QoS} information from previously traversed cells. We demonstrate the application of \ac{LLS} for common channel-aware, as well as channel and queue-aware schedulers. The developed long-term schedulers also provide a controllable trade-off between emphasizing the immediate user \ac{QoS} or the long-term measures. Our simulation results show high gains in long-term \ac{QoS} without sacrificing short-term user requirements.  Therefore, the proposed scheduling approach improves subscriber satisfaction and increases operational efficiency.
\end{abstract}


\maketitle
\footnote[1]{This is an extended version of the paper that appeared in \cite{abouzeid12:lookback}.}
%\footnotetext[2]{Please ensure that you use the most up to date
%class file, available from the WCM Home Page at\\
%{\tiny\href{http://www3.interscience.wiley.com/journal/76507157/home}{\texttt{http://www3.interscience.wiley.com/journal/76507157/home}}}}
\section{Introduction}
\label{sec:intro}
Mobile traffic is experiencing unprecedented growth rates, driven by the large screens of Smartphones and Tablets coupled with online media streaming \cite{maier10:handheld_traffic_wlan}. At the same time, traffic is becoming more unevenly distributed in space and time \cite{paul11:cellular_traffic_dynamics} with demand
peaks moving across different cells throughout the day. Depending on the current traffic situation, users experience a mix of good and bad service while traversing the network. 
Such varying \acf{QoS} is expected to increase with upcoming small cell deployments \cite{damn11:hetnet_survey}, which will result in users traversing a larger number of cells per session. Furthermore, user session times are also becoming longer, with the growing popularity of social media, video and online gaming. When users spend only a small fraction of their session time in each cell and move across a network with unbalanced load, this will lead to new challenges in long-term \ac{QoS} provisioning. Coping with this spatially varying service quality for mobile users is targeted in this paper.

A closer look at current cellular networks reveals three important characteristics of current schedulers. First, the accurate computation of scheduling weights plays a key role in providing \ac{QoS} guarantees to mobile users. Second, current schedulers compute weights by averaging \ac{PHY} data rate and queue length over time intervals in the order of seconds \cite[Ch. 6]{dahlman11:lteA_book}. Third, this weight computation excludes the user's long-term service experience in previously traversed cells. However, focusing only the current cell and ignoring most of the user's service history will provide unsatisfactory \ac{QoS} to mobile users in the long run.

In this paper, we propose \acf{LLS}. This new scheduling framework is based on two components. First, \acp{BS} aggregate \ac{QoS} indicators, such as a user's \ac{PHY} data rate and queuing state, over tens or hundreds of seconds. Then, the \acp{BS} exchange these values as the users traverse the cellular network. In the second component, the final scheduling weight is computed by combining these long-term measures with the conventional short-term moving average \ac{QoS} indicators of the current cell. By doing so, the scheduler can now account for the users' \ac{QoS} indicators over multiple time scales and multiple cells. In the presented \ac{LLS} framework, different \ac{QoS} indicators can be used for different applications. For example, the long-term average user rate can serve as an indicator of user satisfaction for best effort applications. On the other hand, an estimate of the total amount of re-buffering delays for video streaming gives an indication of the long-term quality of streaming. Therefore, the historical long-term experience from prior cells is incorporated in the scheduling decision.

By enabling resource allocation over multiple cells, LLS reduces the negative effects of uneven traffic distribution without sacrificing spectral efficiency and immediate short-term \ac{QoS} needs of the users. 
For example, if a user with a poor service history enters a new cell, the proposed scheduler will prioritize this user over another incoming user who previously received good service. However, if a user with a poor scheduling history has a bad channel state, other users will be scheduled to efficiently use wireless channel resources.

LLS is a general multi-cell scheduling approach that can be applied to various application specific schedulers. We demonstrate this framework for two practical examples. First, we modify the \ac{PF} scheduler \cite{Viswanath02:opp_beamf_dumb_antennas} to include two measures of user average rates 1) the short-term average computed over a few seconds, and 2) the long-term average computed over multiple cells. While this shows the positive effect of LLS on channel-aware scheduling, we also investigate the potential for channel and queue-aware schedulers. Here, we choose the  \ac{EXP} scheduler as it was shown to have good performance with delay sensitive traffic, by keeping queues stable if it is possible to do so \cite{S.Shakkottai01:EXP}. Both scheduler extensions trade-off QoS indicators at different time scales, without requiring central coordination or excessive signaling. This indicates that LLS can be practically applied in existing cellular networks.

The rest of the paper is organized as follows. \sref{sec:RelWork} reviews the pertinent literature, while \sref{sec:Sys} outlines the system description and provides a background on channel scheduling.
In \sref{sec:TwEffects} we discuss the limitations of single-cell scheduling, and then present the details of the proposed \ac{LLS} scheme in \sref{sec:LBS}. The resulting performance analysis is conducted in \sref{sec:Eval}, followed by our conclusions in \sref{sec:Conc}.

\section{Related Work}
\label{sec:RelWork}
Prior work in \ac{BS} coordination for scheduling has mainly focused on instantaneous cooperation to achieve short-term objectives, i.e. \ac{BS}s coordinate their transmissions periodically to minimize interference, balance load, or perform joint transmissions to a user such as in \acf{CoMP} \cite{CoMP}.

In \cite{P.Frank:CoopInter}, Frank \textit{et al.} propose a scheduling scheme for the 3GPP LTE uplink that accounts for inter-cell interference, and by avoiding high interference situations for users at the cell edge, they improve the average spectral efficiency. Bu \textit{et al.} propose a load balancing scheme  that improves proportional fairness over the network by controlling the association of users among neighboring \ac{BS}s \cite{T.Bu:GPF}. In this work, users are associated to BSs 
according to a network-wide proportional fairness criterion instead of the simple strongest BS signal approach. This scheme is extended in \cite{Son:DynamicAssoc} where partial frequency reuse (an inter-cell interference mitigation mechanism) is jointly optimized with the load balancing in a multi-cell network. %Therefore the authors consider both inter-cell association and intra-cell association in their solution.
More recently, in \cite{Zhou11:Global_PF} the authors consider the case where a user is served by multiple \ac{BS}s simultaneously and propose a scheme that provides instantaneous fairness over the network.

\ac{LLS} differs fundamentally from the above multi-cell coordination approaches. Instead of adjusting scheduling based solely on current user conditions and needs, we propose incorporating the long-term service \emph{history} of the users in prior cells into the scheduling framework. We do so to improve the long-term \ac{QoS} for users as they traverse the network.

In a related, but different approach to providing long-term QoS, predictions of the user \emph{future} rates are incorporated to optimize current resource scheduling \cite{abouzeid13:PRA}. This enables the BS to provide long-term QoS by prioritizing users heading to poor coverage \cite{abouzeid13:PRA},\cite{PRA:PPFmargoliesexploiting}, and prebuffering video content opportunistically \cite{abouzeid14:EE_TVT}. However, in this paper we do not make assumptions on the predictability of the future rates, but rather leverage information of the previous rates allocated to provide long-term service.

\section{System Model and Background}
\label{sec:Sys}
In this section we introduce our system model, performance metrics, as well as the traditional schedulers that we use in the proposed \ac{LLS} framework.

\subsection{Network and Mobility Models}
\label{sec:SystemModel}
We study a network with a \acl{BS} set $\mathcal{M}$ and a user equipment set $\mathcal{N}$. An arbitrary user is denoted by $i \in \mathcal{N}$ and an arbitrary \ac{BS} by $m \in \mathcal{M}$, where the number of \ac{BS} is $|\mathcal{M}|=M$ and the number of users $|\mathcal{N}|=N$.

\begin{figure}
	\centering
	\includegraphics[width=70mm]{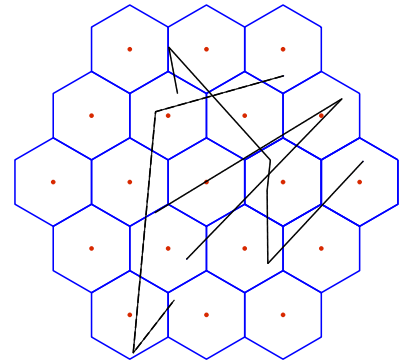}
	\caption{The considered network with 19 cells and 3 sample motion paths.}
	\label{fig:NetworkCluster}
\end{figure}

Each \ac{BS} covers a hexagonal cell. All the users traverse the network according to the \ac{RWP} mobility model with a constant speed $S$, zero pause time between the waypoints, and no wrap-around. Omitting the wrap-around creates a traffic hotspot in the center of the network, which allows us to study an uneven network load distribution. \fref{fig:NetworkCluster} shows the 19 cell network modeled in this paper along with 3 exemplary user motion paths generated using the \ac{RWP} model.

\subsection{Channel and Traffic Models}
We model the wireless downlink as typical for studies on macro-cell \ac{LTE} systems. The path loss is calculated according to \cite{3GPPSpec} as $\textnormal{PL}(d)=128.1+37.6\log_{10} d$, where $d$ is the user-BS distance in km. The \acp{BS} have omnidirectional antennas and a log-normal distribution with a variance of 8\,dB accounts for slow fading. Fast fading is modeled as i.i.d. Rayleigh-fading, and link adaptation is modeled using Shannon's equation where the SNR is clipped at 20\,dB to account for a maximum modulation order of 64 \ac{QAM}.

\begin{figure*}
	\noindent
	\centering
	\includegraphics[width=110mm]{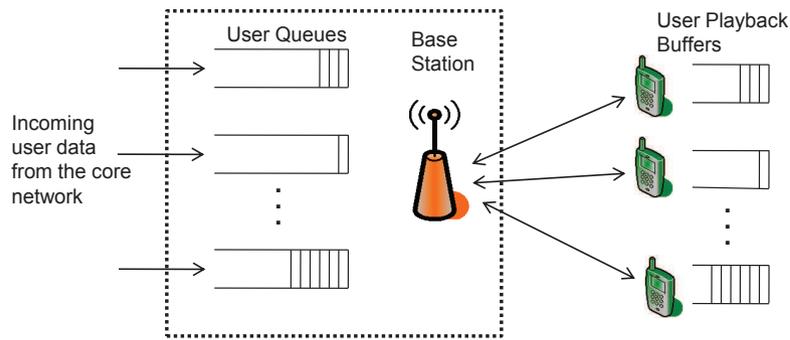}
	\caption{Studied system with the traffic model for buffered media streaming.}
	\label{fig:SystemModel}
\end{figure*}

Two traffic models are considered in this paper. The first is full buffer, meaning that each user has download traffic at any point in time. This model is used to compare the performance bounds for schedulers that are unaware of BS queues. The second traffic model includes user queues at the BS as shown in \fref{fig:SystemModel}. For each of these queues, packets arrive at a constant rate $\lambda_i$. This model is used to evaluate queue-aware schedulers and represents non-real time video streaming traffic arriving at the BS with a constant bit rate. 

To account for the buffering of such traffic, we also consider playback buffers at the user terminals. Here, the video stream will play at a constant rate $R_\textnormal{Stream}$ when the buffer is sufficiently filled. The video stream will freeze (or stall) when the buffer becomes empty if the user has not been scheduled sufficiently. The stream will remain frozen until the playback buffer is re-filled to the playback threshold, which corresponds to the behavior of modern stream protocols such as \ac{HLS} \cite{pantos11:hls_draft_ietf}.
\subsection{Channel Schedulers}
In this paper we focus on downlink scheduling where the \ac{BS}
makes user scheduling decisions every time slot. We model a discrete time slotted channel where each slot is referred to as a \ac{TTI}. Users report their \ac{CQI} (represented as the instantaneous achievable rate $r_i(t)$) every \ac{TTI}, which  \acp{BS} use to make scheduling decisions in the upcoming \ac{TTI}. In the following we review the considered downlink schedulers.

\subsubsection{Max-rate Scheduling}
The \ac{MR} rule schedules the user with the highest instantaneous \ac{CQI}, as observed from the previous \ac{TTI} \cite{Viswanath02:opp_beamf_dumb_antennas}. This maximizes the sum throughput of the network but makes no effort to serve users fairly.

\subsubsection{Proportional Fair Scheduling}
The \acf{PF} scheduling rule \cite{Kelly1997} aims for high throughput while maintaining fairness among the users. The intuition of the algorithm is to schedule users when they are at their peak rates \textit{relative} to their own average rates. At any \ac{TTI} $t$, PF schedules the user $i^*=\arg \max_{\forall i \in \mathcal{N}} w_i(t)$ where the user weights are calculated $\forall i \in \mathcal{N}: w_i(t)=r_i(t)/R_i(t)$. Here, $r_i(t)$ refers to the instantaneous data rate in the last time slot while $R_i(t)$ is the moving average of the data rate, computed as \cite{Viswanath02:opp_beamf_dumb_antennas}
\begin{equation}
	\label{eq:PFAv}
	R_i(t+1) = \frac{1}{W} r_i(t) p_i(t) + \left(1-\frac{1}{W}\right) R_i(t).
\end{equation}
Here, $p_i(t)$ is a binary variable, which is equal to $1$ when user $i$ is scheduled at slot $t$ and equal to $0$ otherwise. The parameter $W$ denotes the time window, over which the moving average is computed. If a user remains in good channel conditions, it's average rate $R_i(t)$ will also be high. Therefore, by scheduling according to $r_i(t)/R_i(t)$, the \ac{BS} in effect compares the current user achievable rate to the user allocation \textit{history}, and selects the user with the highest relative measure. This is equivalent to scheduling users when they are at their own channel peaks. Note that the size of $W$ defines the duration over which the user allocation history is computed, and is therefore tied to the latency of the application. A smaller value of $W$ will direct the scheduler to make frequent user allocations to ensure that their average rate $R_i(t)$ does not fall to zero during $W$. Conversely, a large $W$ enables the scheduler to wait longer before scheduling a user when its channel hits a very high peak. Such a delay tolerance will result in an increased system throughput.

\subsubsection{Exponential Scheduler}
\label{sec:EXP}
The \acf{EXP} \cite{S.Shakkottai01:EXP} scheduler is \textit{queue-aware} in addition to being channel-aware. By incorporating information of the user queue lengths, it can reduce the delay of buffered data in user queues. In the \ac{EXP} scheduling rule, when a user queue gets large relative to the other users' queues, its scheduling priority is increased exponentially. Put formally, this scheduling rule chooses user
\begin{equation}
	\label{eq:EXP}
	i^*=\arg \max_{\forall i \in \mathcal{N}} 
	\frac{r_i(t)}{R_i(t)} \exp{\frac{a_i q_i(t) - \frac{1}{N}\sum_{i=1}^{N} a_i q_i(t)}{1 + \sqrt{\frac{1}{N}\sum_{i=1}^{N} a_i q_i(t)} } }	
\end{equation}
where $q_i(t)$ is the queue length for user $i$ at time slot $t$. The parameter $a_i$ allows the scheduler to prioritize certain user queues over others. It controls the strictness of the scheduler in responding to growing queue lengths of each user.

\subsection{Performance Metrics}
\noindent We study the following performance metrics:
\begin{itemize}
\item $T_{\textnormal{Net}}$: the average network throughput which is measured during the downlink as
 the sum of the average data rate taken over all users of the network.
 \item $J_{\textnormal{Net}}$: Jain's fairness index for user throughput and is computed as $(\sum_{i=1}^N T_i)^2/N \sum_{i=1}^N T_i^2$ where $T_i$ is the average throughput for user $i$ over the time of interest. We use this metric to compute the long-term throughput fairness of the network.
 \item $T_{\textnormal{Slot}}^{10\%}$: the average 10th percentile user throughput w.r.t. time. We developed this metric to quantify the users' rate starvation. For each user, we first compute the received average throughput during a 'bin' of one second. For a time duration of several hundred seconds, this results in a vector of values. Then, we compute the 10th percentile throughput of this vector. If the value is low this indicates that several time bins had poor throughput, and thus, that user was starved. $T_{\textnormal{Slot}}^{10\%}$ gives the average 10th percentile vector values obtained from all the users.
 \item $F^{\tn{LT}}$: the average long-term amount of video freezing experienced by all users in the network expressed as a percentage of the playback duration.
 \item $J^F_{\tn{Net}}$: Jain's fairness index of user video freezing.
 \end{itemize}

\section{Limitations of Single-cell Scheduling}
\label{sec:TwEffects}
In current networks, the average user rate ${R}_i(t)$ in \eref{eq:PFAv} is computed at each \ac{BS} independently. This works well if $W$ is small. However, in the case of a large $W$, when a user with an ongoing session moves from one cell to another, ${R}_i(t-1)$ from the previous cell is unknown to the current cell. This means that the current cell will restart the computation of \eref{eq:PFAv} without the users past history, hence introducing errors in the calculation of ${R}_i(t)$. 
This poses a limitation for any scheduling rule that relies on the historical average user rate (or in general, any user quality metric computed over a time duration). For example, in \ac{PF} this results in throughput and fairness losses for large window sizes, which we discuss in the following. 

\begin{figure}
	\noindent \centering
	\includegraphics[width=70mm]{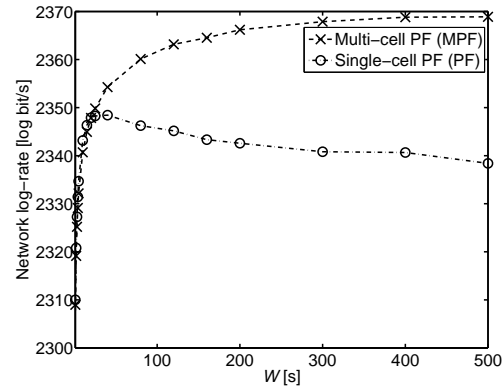}
	\caption{Effect of the averaging window size $W$ on $R_\textnormal{log}^\textnormal{net}$ for single-cell and multi-cell PF. $N=250$ users, $S=40$\,km/h.}
	\label{fig:LogSum}
\end{figure}

A scheduler is said to be proportionally fair if it maximizes the sum of the logarithmic rate of all the users \cite[App. A]{Viswanath02:opp_beamf_dumb_antennas}
 \begin{equation}
	\label{eq:LogSumR}
	R_\textnormal{log}^\textnormal{net} = \sum_{i=1}^{N} \log \bar R_i,
\end{equation}
for an asymptotically large $W$. Here $\bar R_i$ is the exact average user rate during $W$
\begin{equation}
	\label{eq:UserAv}
	\bar R_i = \frac{1}{W} \sum_{t=1}^{W} r_i(t) p_i(t).
\end{equation}
%the \ac{PF} property from . A scheduler fulfills this property if it 
In \fref{fig:LogSum} we show the effect of increasing $W$ on $R_\tn{log}^\tn{net}$ based on a simulation of the network in \fref{fig:NetworkCluster}. The single-cell PF scheduler is one where the scheduler computes \eref{eq:PFAv} without knowledge from previous BSs, whereas in the multi-cell PF, BSs have complete user rate information. While the multi-cell PF maintains a rate increase with $W$, the $R_\tn{log}^\tn{net}$ of the single-cell PF 
 starts decreasing after a $W$ of $20$\,s.
At $W>20$\,s the rate information in one cell cannot accurately represent the real average rate a user has experienced. 
Single-cell PF clearly does not maximize $R_\textnormal{log}^\textnormal{net}$ for increasing $W$, and therefore violates the \ac{PF} property \cite[App. A]{Viswanath02:opp_beamf_dumb_antennas}. This means that it cannot provide proportional fairness over multiple-cells.

\begin{figure}[!b]
	\centering
	\includegraphics[width=60mm]{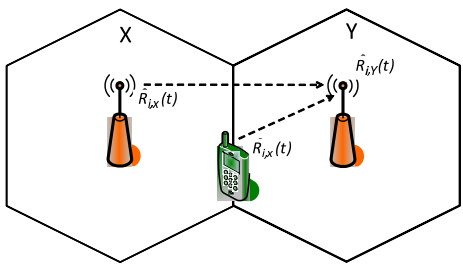}
	\caption{As user $i$ enters cell $Y$, $\bar{R}_{i,x}(t)$ is transmitted to $\textnormal{BS}_Y$ from either $\textnormal{BS}_X$ or the user terminal. $\bar{R}_{i,x}(t)$ represents the average rate the user received in cell $X$.}
	\label{fig:MPF}
\end{figure}
\begin{figure*}
	\noindent
	\centering
	\includegraphics[width=100mm]{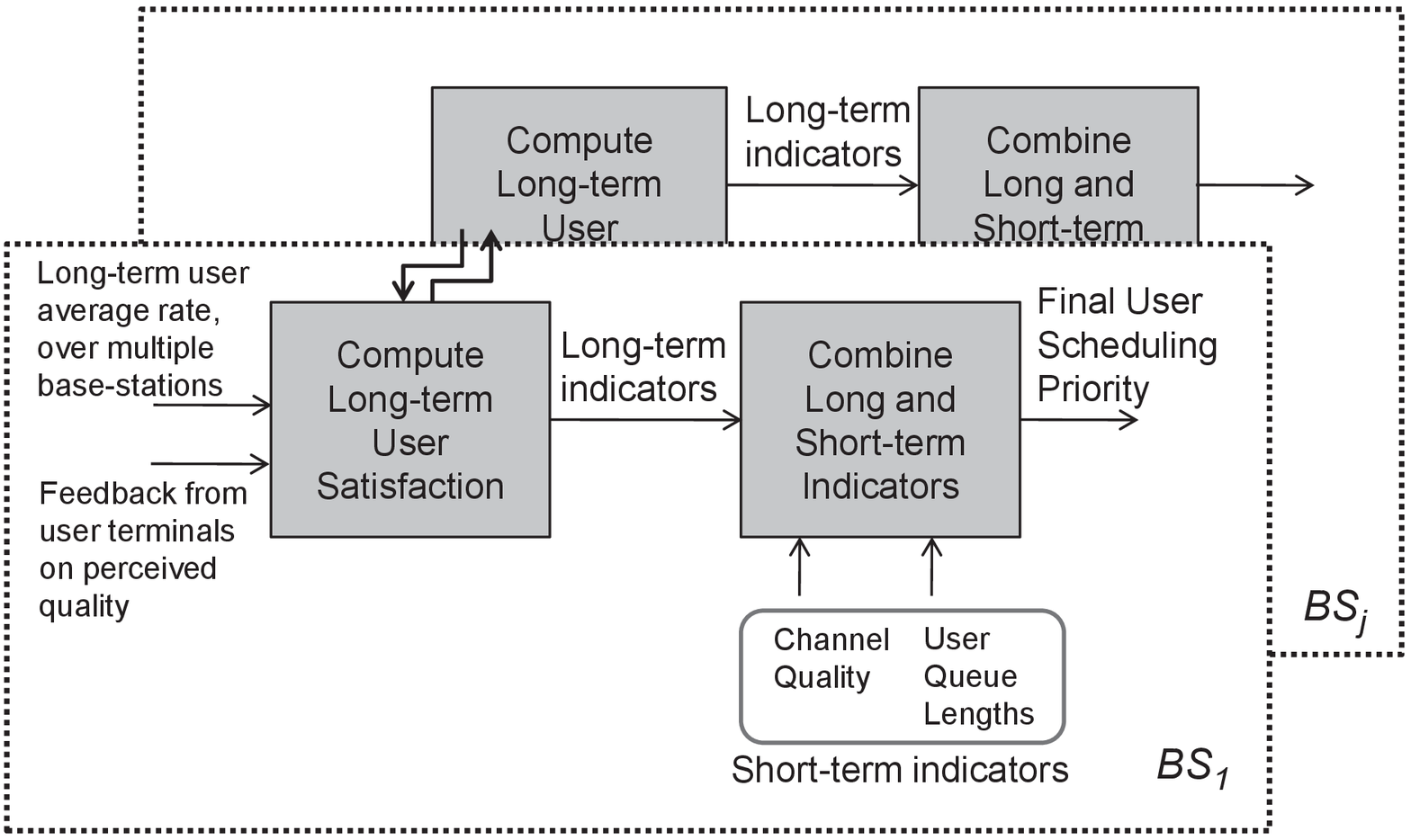}
	\caption{The proposed \acf{LLS} framework.}
	\label{fig:LLBScheduler}
\end{figure*}
From an implementation point of view, multi-cell PF can be achieved in two ways as shown in \fref{fig:MPF}. First, $\tn{BS}_X$ can signal the average user rate $\hat{R}_{i,x}(t)$ to $\tn{BS}_Y$ via interfaces such as the X2-interface in \ac{LTE} \cite{dahlman11:lteA_book}. If such signaling is not supported, the user handheld can receive the average rate from $\textnormal{BS}_X$ (or compute it itself) and signal it to $\textnormal{BS}_Y$ during handover via the air link. Note that either of these options requires only an exchange of one average rate value per user, per handover. As a typical \ac{CQI} resolution is 4 bits in LTE \cite{hsdpa11:phy_spec_cqi_res}, this adds only an insignificant signaling effort.

\section{Long-term Lookback Scheduling}
\label{sec:LBS}

Conventional single-cell schedulers are based on metrics computed at the scheduling \ac{BS} with no regard to the users' scheduling history in previously traversed cells. In \sref{sec:TwEffects} we discussed how scheduling can be extended to the multi-cell case by signaling historical rates received (or any other \ac{QoS} indicator used in scheduling) during hand-over. However, basing the scheduling on these long-term rates computed over multiple cells will not guarantee that the short-term user requirements are satisfied. Therefore, we argue that both the long and short-term user rates and \ac{QoS} indicators should be combined in a single scheduling framework. 

\subsection{Scheduling Framework}
The proposed \acf{LLS} framework is shown in \fref{fig:LLBScheduler}. Like existing schedulers, it uses short-term QoS indicators such as instantaneous channel gain and user queue lengths, which are evaluated at each \ac{BS}. Our framework introduces a module to compute long-term user satisfaction, which is based on the average rate a user received over multiple cells, or more application specific \ac{QoS} satisfaction indicators that may be fed back directly from user terminals. These long-term measures are computed over tens or hundreds of seconds and exchanged between \ac{BS}s when users are handed over as shown in \fref{fig:LLBScheduler}.
As previously discussed, such an exchange may be possible via the X2-interface in LTE \cite{dahlman11:lteA_book} and will not add significant overhead.
In the following, we present the application of the \ac{LLS} framework to incorporate long-term indicators in the \ac{PF} and \ac{EXP} schedulers.

\begin{figure*}
    \centering
	\subfigure[Exponential utilities: $\exp(\alpha/x)$.]{
	\includegraphics[width=75mm]{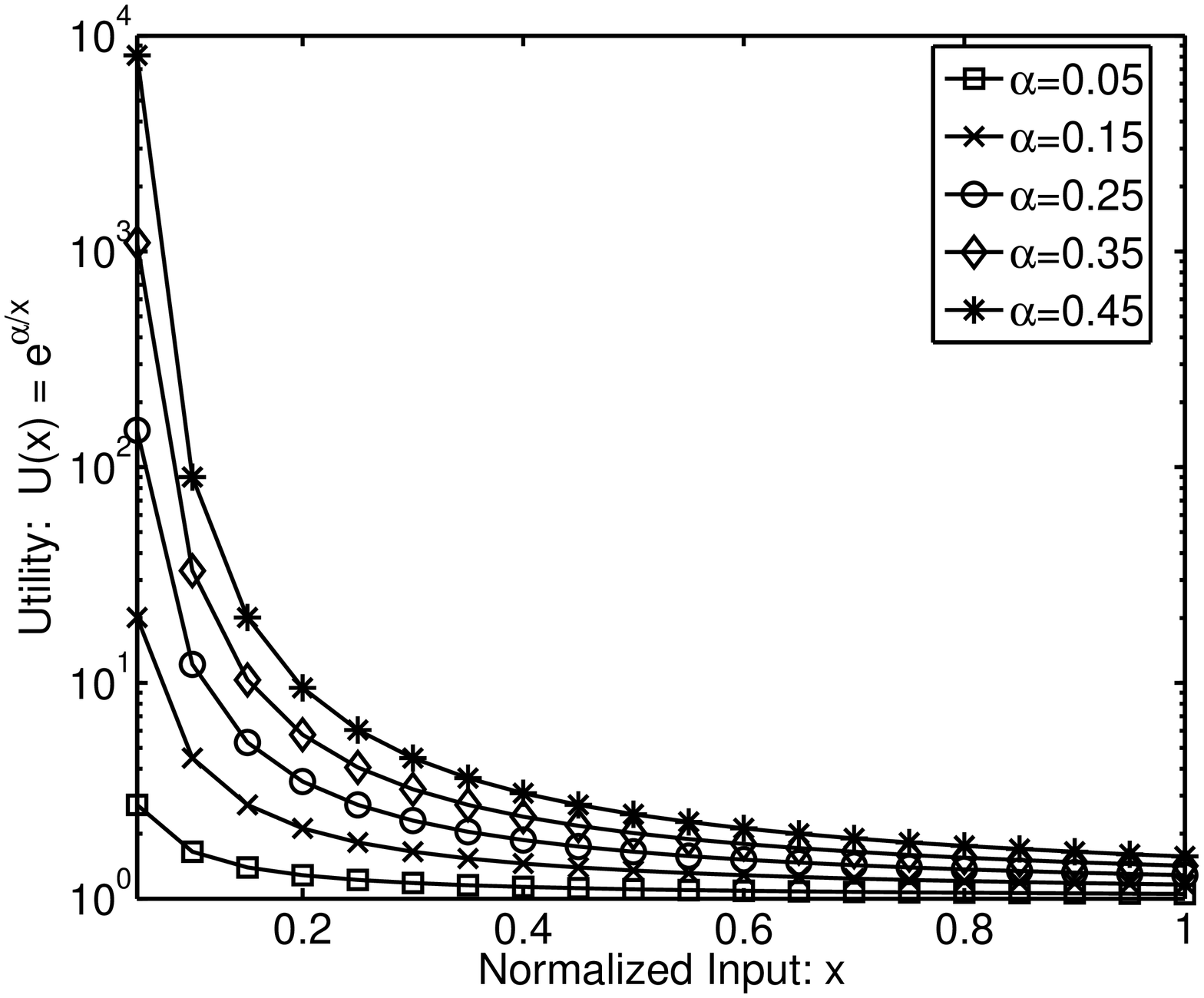}\label{fig:Utility_Exp}}
	\subfigure[Sigmoid utilities: $1-\exp(-c(x-\beta))$.]{
	\includegraphics[width=75mm]{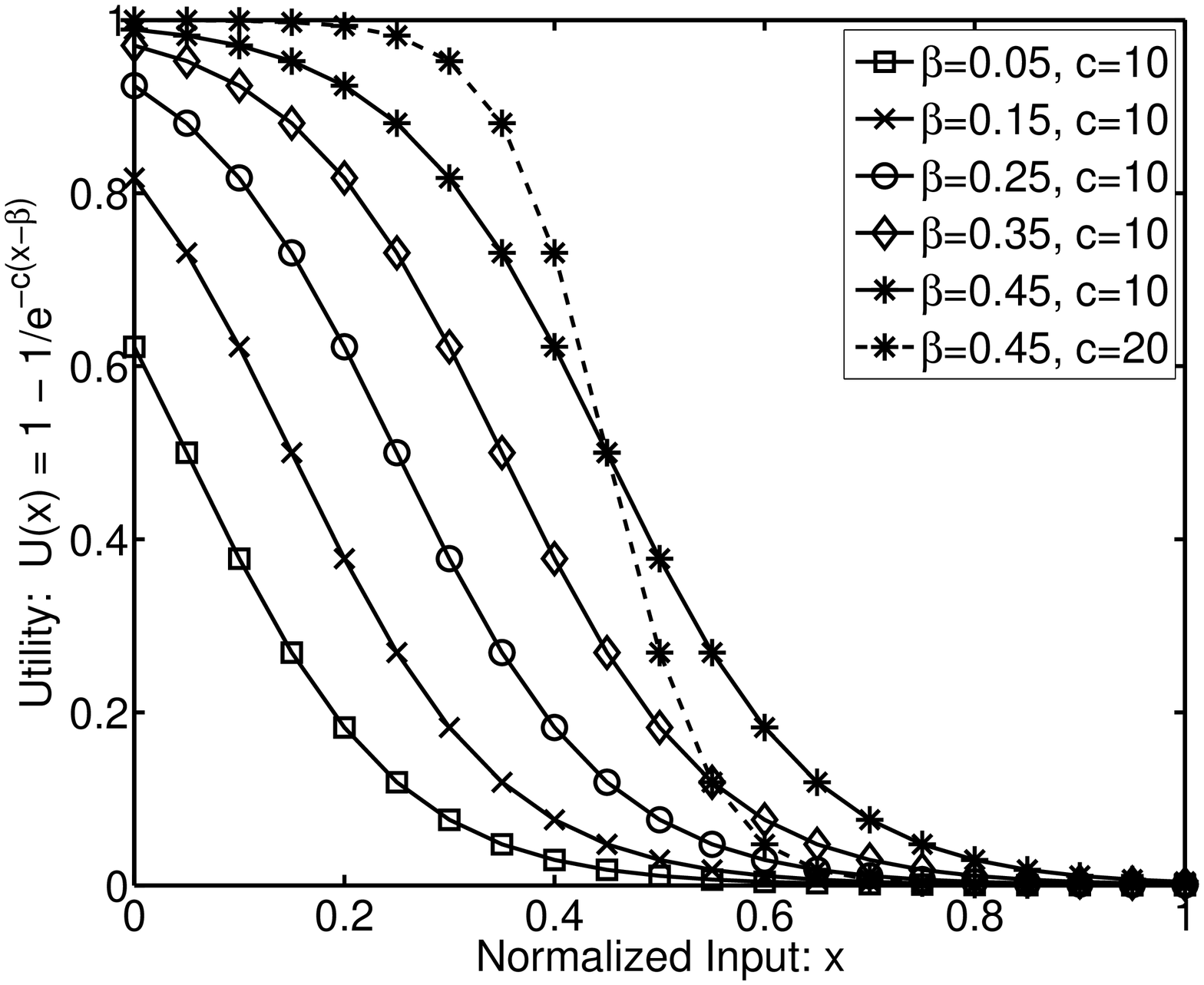}\label{fig:Utility_Sig}}
	\caption{Effects of parameter choices on utility functions.}
\label{fig:Utility_Design}
\end{figure*}

\begin{figure*}
    \centering
	\subfigure[LL-PF-Exp: Exponential short-term utility, $\alpha=0.2$. Rates are normalized.]{
	\includegraphics[width=75mm]{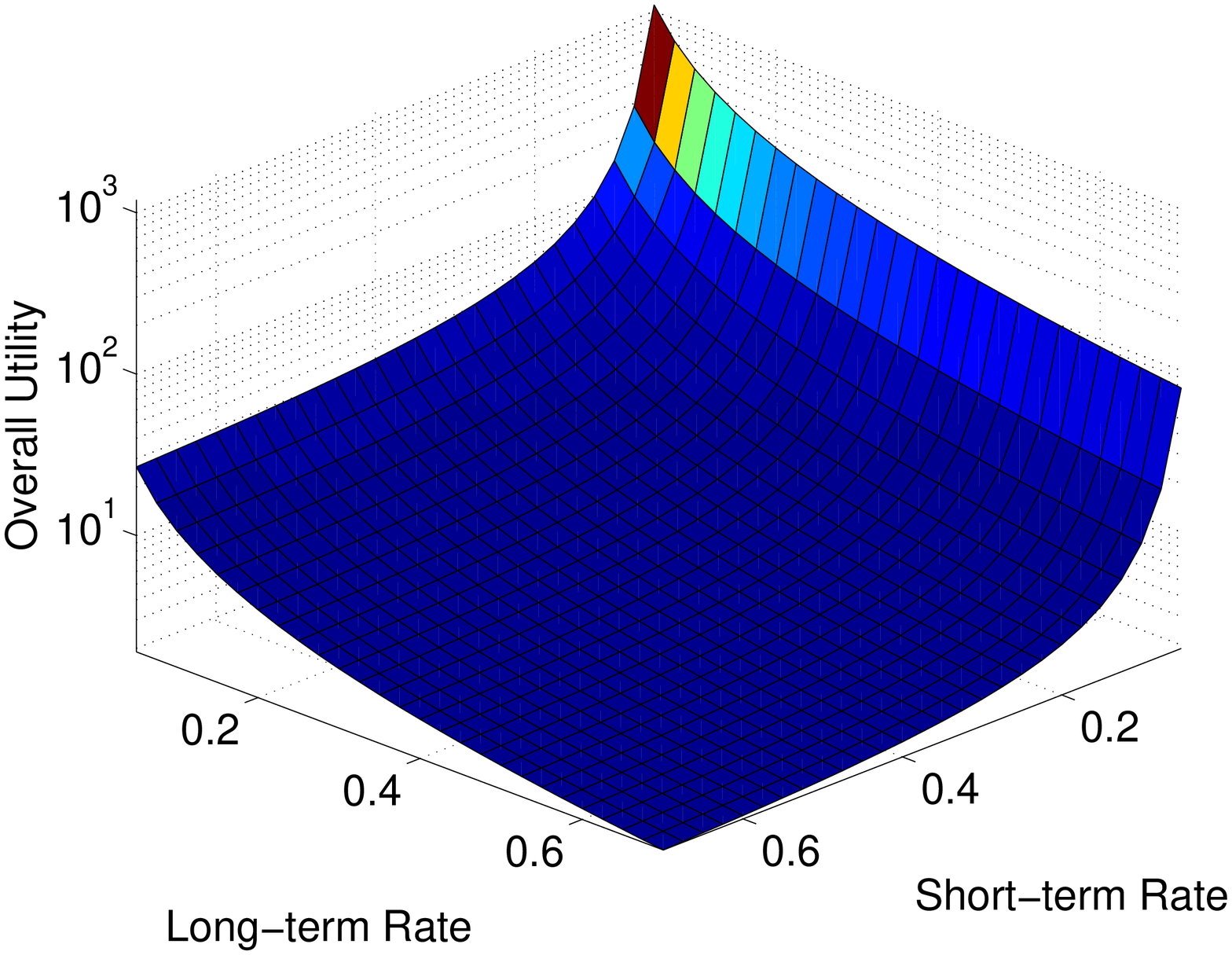}\label{fig:PF_Exp}}
	\subfigure[LL-PF-Sig: Sigmoid short-term utility, $\beta=0.5, c=10$. Rates are normalized.]{
	\includegraphics[width=75mm]{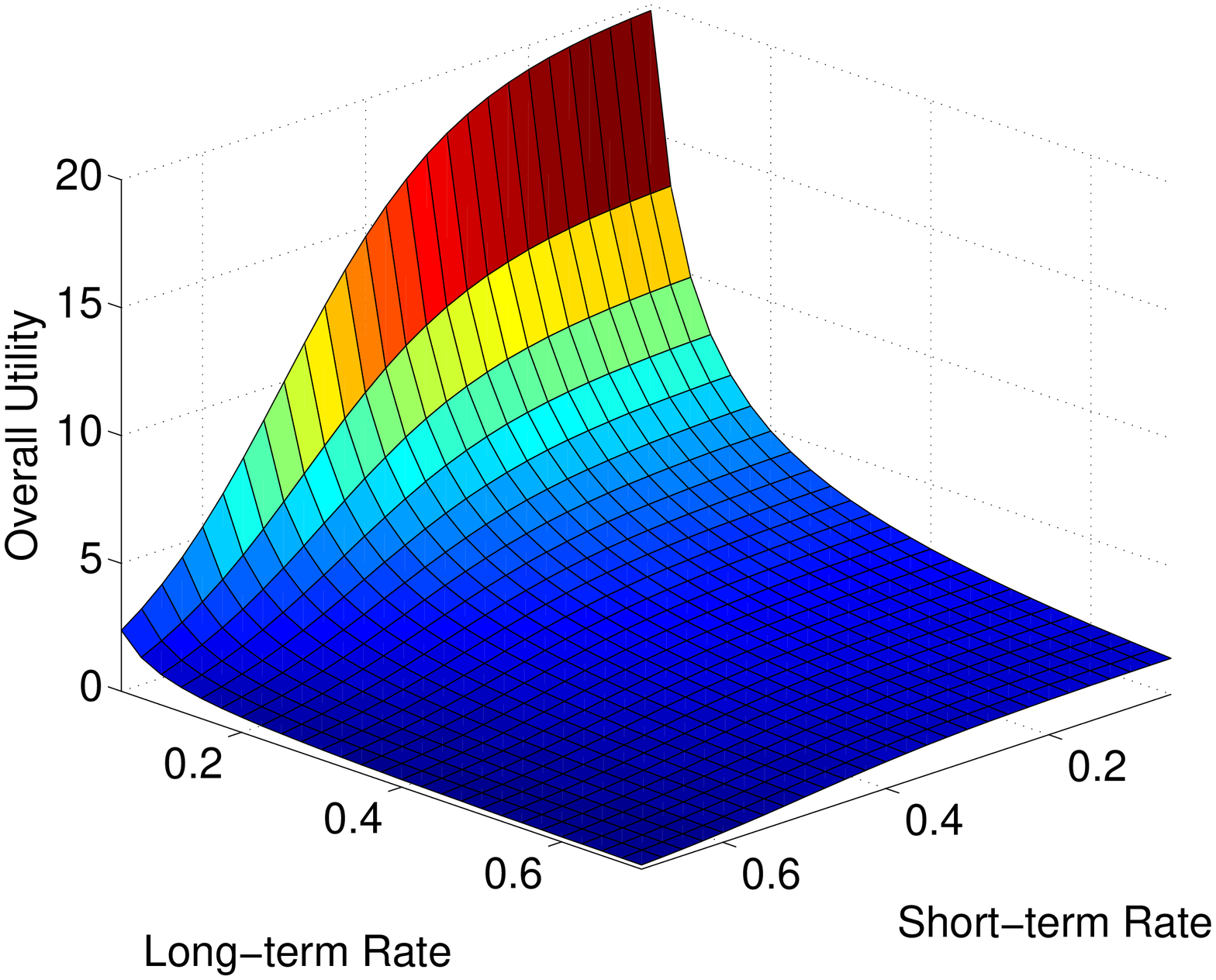}\label{fig:PF_Sig}}
	\caption{Combined short and long-term rate utilities.}
\label{fig:Utility_Design}
\end{figure*}
\subsection{Utility Selection}
An important design choice in the \ac{LLS} framework is the definition of utility functions for the short and long-term indicators. The shape of the utility functions will determine how the scheduler responds to changes in the indicators. For example, 
\fref{fig:Utility_Exp} illustrates an exponential utility of the form $U(x)=\exp(\alpha/x)$, where $x\in[0.05,1]$ to avoid an unbounded response. Let us assume that $x$ denotes the average user rate normalized w.r.t. the user with the highest rate at the current TTI. Then, such a utility function will prioritize users with very low rates. The degree of this prioritization is a function of $\alpha$ as shown in \fref{fig:Utility_Exp}. 

Another useful choice for utilities is the 'S-shape' or sigmoid function illustrated in \fref{fig:Utility_Sig}. As shown, changes in the input $x$ influence the utility exponentially but saturate at either extreme end of $x$. The plotted function is 
$U(x)=1-\exp(-c(x-\beta))$, where $\beta$ controls the inflection point (where the slope is at a maximum magnitude) and $c$ controls the steepness of the curve. Increasing $\beta$ will result in earlier response to a decreasing QoS indicator, thereby prioritizing the user as soon as it falls below the acceptable value. A higher slope $c$ will suddenly increase the utility as the value of $x$ decreases, with the change happening in the vicinity of the value of $\beta$ as shown in \fref{fig:Utility_Sig}. Therefore, the sigmoid function offers a wide range of response options which we discuss in \sref{sec:Eval}.

\subsection{\acf{LL-PF} Scheduler}
The \ac{LL-PF} scheduler is proposed to maintain long-term fairness between users, while simultaneously providing the flexibility to serve a variety of delay sensitive applications.
% achieve a mix of short and long-term fairness over multiple cells. It provides the flexibility to serve a variety of applications while maintaining long-term fairness between users. 
This scheduler is only channel aware, and uses the average user-rate for scheduling decisions. As opposed to traditional rate-based schedulers, \ac{LL-PF} computes the average user rate over both short and long durations. We first present LL-PF-Exp, where these two satisfaction indicators are combined using an exponential utility for the short-term user rate, while the long-term rate follows a $1/x$ utility. In LL-PF-Exp, users are scheduled according to
\begin{equation}
	\label{eq:PF-LT}
	i^*=\arg \max_{\forall i \in \mathcal{N}} 
	\frac{r_i(t)}{R_{i,m(t)}^\textnormal{LT}(t)} \exp{\frac{\alpha_i}{R_i^\textnormal{norm}(t)}}   	
\end{equation}
where $R_i^\textnormal{norm}$ is the short-term average rate (so \eref{eq:PFAv} is computed with a $W$ of a few seconds), normalized by the highest user rate value. Therefore, $R_i^\textnormal{norm} \in [0,1]$. By choosing the exponential function for short-term utility, the user priority will increase exponentially as $R_i^\textnormal{norm}(t)$ decreases. Thereby, we ensure that users do not starve. The computation of $R_{i,m(t)}^\textnormal{LT}(t)$ in \eref{eq:PF-LT} is according to 
%\begin{fleqn}
\begin{equation}
\begin{split}
\label{eq:MPF}
	R_{i,m(t)}^\textnormal{LT}(t+1) =
	&\frac{1}{W} r_{i,m(t)}(t) p_i(t)\\ &+
	 \left(1-\frac{1}{W}\right) R_{i,m(t-1)}^\textnormal{LT}(t),
\end{split}
\end{equation}
%\end{fleqn}
which denotes the long-term average user rate, over several base-stations, computed at \ac{BS} $m$. Therein, ${m(t), m(t-1) \in \mathcal{M}}$ are the respective BS indices in the current and previous time slot. If a user changes the cell, $m(t) \neq m(t-1)$ but the average is still computed as \ac{BS}s exchange the value of $R_{i,m(t)}(t)$ during handover. Note that the time window $W$ is significantly longer than in the computation of the short-term moving average. Choosing a large value for  $W$, provides user fairness over a longer duration. 

\begin{figure*}
	\noindent
	\centering
	\includegraphics[width=110mm]{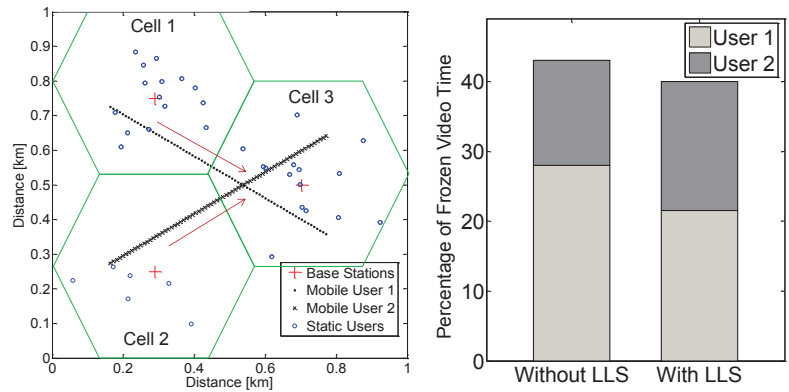}
	\caption{Long-term Lookback Scheduling improves video quality experience and provides more fair service over multiple cells.}
	\label{fig:SyntheticResults}
\end{figure*}
The parameter $\alpha$ determines the rate at which the exponential factor of the short-term user rate increases, and can have different values for each user. Increasing $\alpha$ will make the scheduler biased towards providing short-term fairness, and a value of $0$ will make the scheduler a purely long-term multi-cell proportional fair scheduler, which we presented in \cite{Abou-zeid12:MPFS}. \fref{fig:PF_Exp} illustrates the overall utility of LL-PF-Exp, and how it is affected by the changes in both long and short-term user rates, for $\alpha=0.2$. Although the value of $\alpha$ depends on the application preference, we will see in \sref{sec:Eval} that LL-PF-Exp maintains a higher level of fairness than traditional PF for any level of $\alpha$.

With a similar intuition, we present the LL-PF-Sig scheduler, where the short-term user rates are dependent on a sigmoid utility. This offers a broader range of potential scheduler behavior. In this case, the scheduler selects the user $i^*$ that satisfies:
\begin{equation}
	\label{eq:PF-LT-Sig}
	i^*=\arg \max_{\forall i \in \mathcal{N}} 
	\frac{r_i(t)}{R_{i,m(t)}^\textnormal{LT}(t)} 
	\bigl(1-\exp(-c(R_i^\tn{norm}(t)-\beta))\bigr).
	%\exp{\frac{\alpha_i}{R_i^\textnormal{norm}(t)}}   	
\end{equation}
\fref{fig:PF_Sig} illustrates a sample overall utility of LL-PF-Sig. We can see that compared to \fref{fig:PF_Exp}, the short-term rate will have an immediate affect on the overall utility, and with such a parameterization, user short-term starvation will be prevented.

\subsection{\acf{LL-EXP} Scheduler}
The \ac{LL-EXP} extends the channel and queue-aware scheduler from \sref{sec:EXP} to include long-term \ac{QoS} indicators. 
In the first embodiment of this scheduler we keep the instantaneous user queue size as a short-term scheduling indicator (to accomodate delay sensitive traffic) but replace the average user rate $R_i(t)$ in \eref{eq:EXP} with the long-term average rate $R_{i,m(t)}^\textnormal{LT}(t)$. Depending on the user's trajectory, these long-term averages may be computed over several \acp{BS}. 
Therefore, \ac{LL-EXP} algorithm schedules user $i^*$ that satisfies:
\begin{equation}
	\label{eq:EXP-LT}
	\arg \max_{\forall i \in \mathcal{N}} 
	\frac{r_i(t)}{R_{i,m(t)}^\textnormal{LT}(t)} \exp{\frac{a_i q_i(t) - \frac{1}{N}\sum_{i=1}^{N} a_i q_i(t)}{1 + \sqrt{\frac{1}{N}\sum_{i=1}^{N} a_i q_i(t)} } }.	
\end{equation}

Note that by relaxing the duration over which the user average is computed, the scheduler can be more opportunistic in serving users that have a high instantaneous rate $r_i(t)$. Results in \sref{sec:Eval} indicate that this reduces the likelihood of video freezing experienced by users. 

\subsubsection{LL-EXP with Video Freezing Feedback}
In this extension of the \ac{EXP} scheduler we directly consider the amount of long-term video freezing $F_i^\textnormal{LT}(t)$ experienced by the users. The idea is to multiply the scheduling utility by the historical average amount of video freezing experienced. This will decrease the priority of users will low freezing and attempt to limit the total amount of video stalling a user will experience throughout a session. As shown in the results, this also provides a high degree of fairness in video freezing. It is assumed that ${F}_i^\textnormal{LT}(t)$ is either fed back by the users to the BS or estimated by \ac{QoS} monitoring tools at the BS. A user $i^*$ is scheduled that satisfies:
\begin{equation}
	\label{eq:EXP-F}
	\arg \max_{\forall i \in \mathcal{N}} 
	r_i(t) {F}_i^\textnormal{LT}(t) \exp{\frac{a_i q_i(t) - \frac{1}{N}\sum_{i=1}^{N} a_i q_i(t)}{1 + \sqrt{\frac{1}{N}\sum_{i=1}^{N} a_i q_i(t)} } } 	
\end{equation}
In what follows we shall refer to this as the LL-Exp-Freeze scheduler.

\section{Performance Evaluation}
\label{sec:Eval}

In this section we first illustrate our scheduling framework for a very simple scenario. Then we study a more general simulation set-up and discuss the performances of each of the presented long-term lookback schedulers.

\subsection{Simple Scenario}
Consider the scenario of \fref{fig:SyntheticResults} where two users watching a video stream are moving towards $\textnormal{Cell 3}$. User 1 is arriving from a congested cell and suffered excessive video freezing, whereas User 2 is coming from a sparsely populated cell and experienced better playback. As Cell 3 is also congested, both users will now be subject to video freezing. In the traditional scheduling approaches both users will suffer equally on arrival at $\textnormal{BS}_3$. This, however, can be changed if $\textnormal{BS}_3$ is made aware of the freezing history of User 1 in its previous cell. Now, $\textnormal{BS}_3$ can increase the scheduling weight of this user to increase its QoS.

A sample result from applying the proposed LL-EXP-Freeze scheduler is shown by the bar plot in \fref{fig:SyntheticResults}. Here, we can see that total freezing for User 1 is reduced from 28\% to 21\% compared to the case of scheduling without \ac{LLS} (i.e. no service history from prior cells). User 2 on the other hand suffers slightly more freezing in \ac{LLS} than in the case without \ac{LLS}. This indicates that, with information from previous cells, \ac{LLS} can allocate resources to provide a more fair video experience to the users. The total amount of freezing for both users has also been reduced. 

\subsection{Simulation Set-up}
We evaluate the schedulers in the 19 cell network of \fref{fig:NetworkCluster} with an inter-BS distance $D=1$\,km, and a BS transmit power of 40\,W for the downlink. The center carrier frequency is 2\,GHz, while we choose a bandwidth of $10$\,MHz for the full buffer traffic scenario and  $5$\,MHz for the buffered video streaming traffic (assuming the remaining $5$\,MHz are used for other services). User speed $S=10$\,m/s. Simulations run for $500$\,s simulated time, with a $200$\,s warm-up period. 

\begin{figure*}
    \centering
	\subfigure[Average network throughput $T_{\textnormal{Net}}$]{
	\includegraphics[width=75mm]{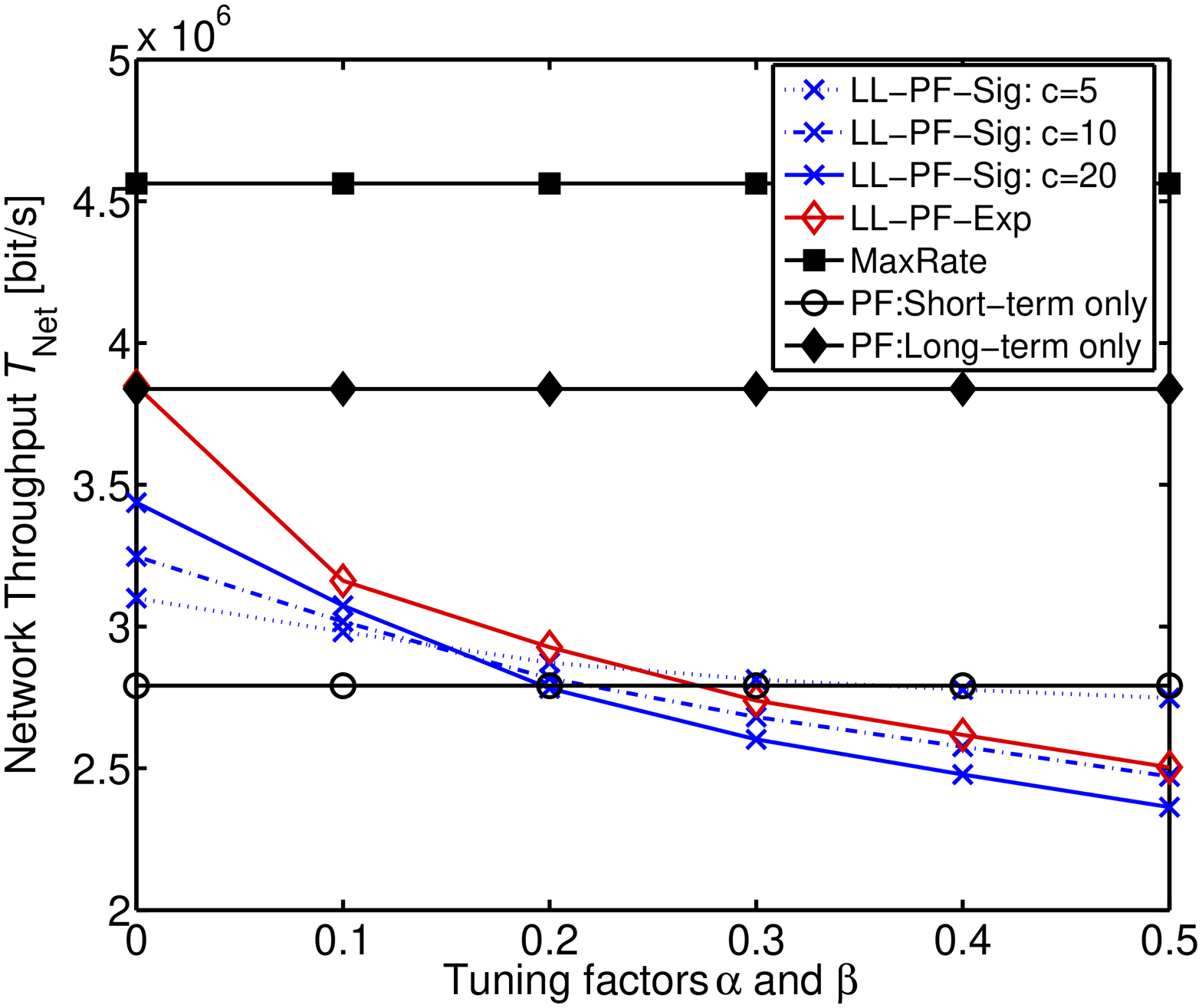}\label{fig:TvsAlpha}}
%	\quad
	\subfigure[10th \% slot throughput $T^{\textnormal{slot}}_{10\%}$.]{
	\includegraphics[width=75mm]{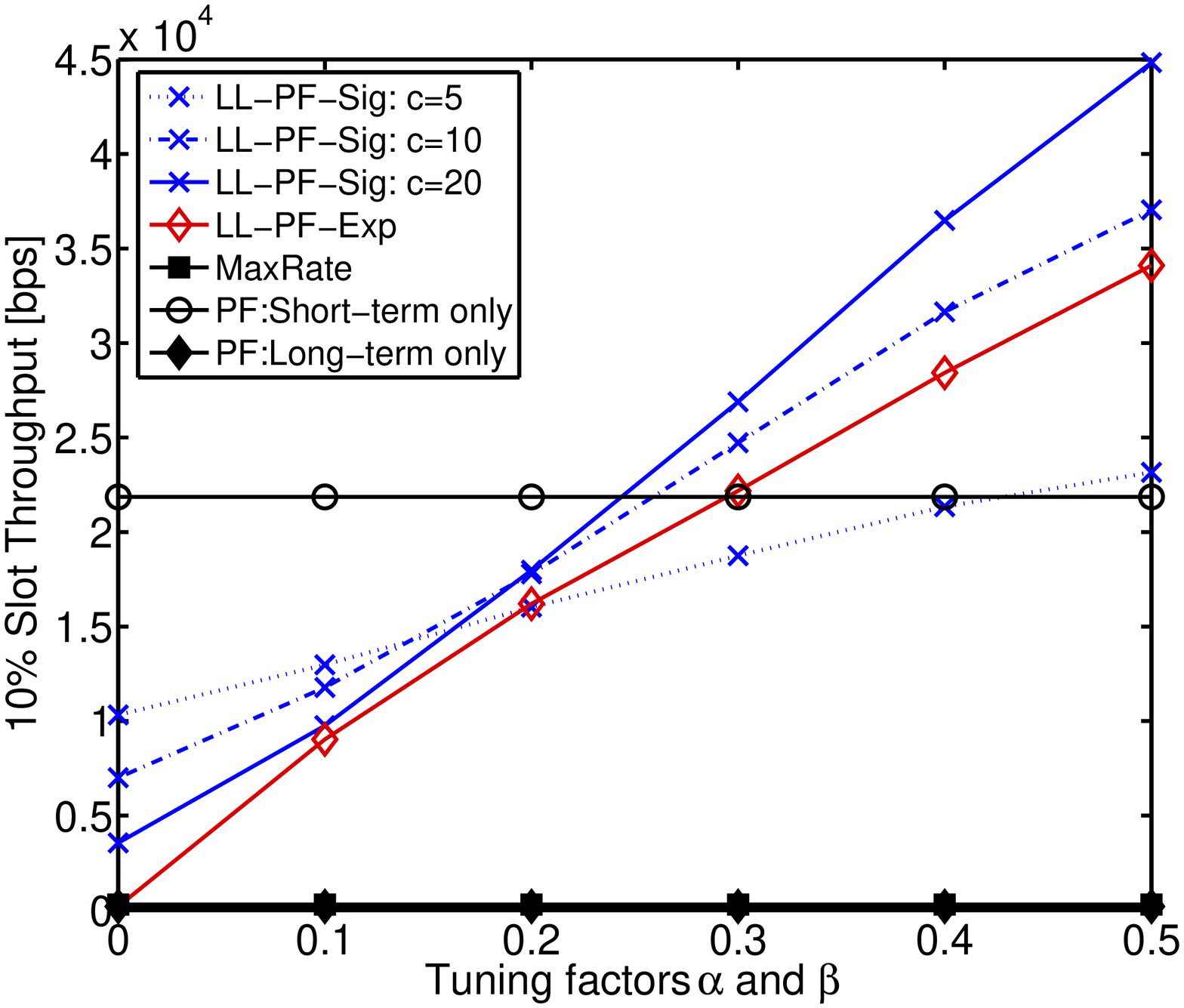}\label{fig:SlotvsAlpha}}
	\caption{Effect of $\alpha$ and $\beta$ on the network throughput $T_{\textnormal{Net}}$, and 10th \% slot throughput $T^{10\%}_{\textnormal{Slot}}$, at a user speed of $S=40$\,km/h, and $N=200$. The x-axis reads values of $\alpha$ for LL-PF-Exp, and reads the values of $\beta$ for LL-PF-Sig.}
	\label{fig:Alpha_PF-LT}
\end{figure*}

\subsection{Performance of \ac{LL-PF}}
For this study we assume that all the users have the same priority value of $\alpha_i$, and set the time window $W$ to  $1$\,s for the short-term average, and to $300$\,s for the long-term average. We use a full buffer traffic model to study the schedulers' highest performance in a saturated network.

\begin{figure}
    \noindent
    \centering
	\includegraphics[width=75mm]{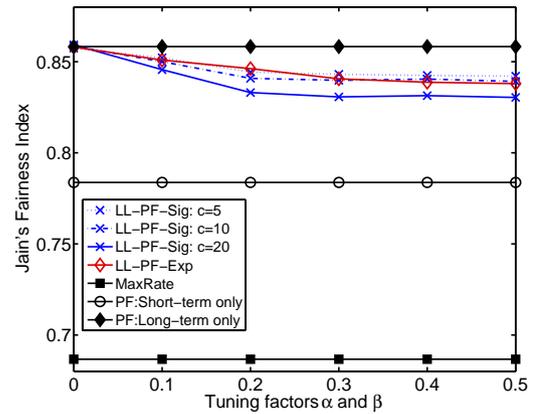}
	\caption{Effect of $\alpha$ and $\beta$ on  Jain's fairness index of the long-term user rates, at a user speed of $S=40$\,km/h, and $N=200$.}
	\label{fig:JvsAlpha}
\end{figure}

\fref{fig:TvsAlpha} shows the throughput performance of \ac{LL-PF} for different values of $\alpha$. When $\alpha=0$, the {LL-PF-Exp} throughput is much larger than \ac{PF}-Short-term and approaches the throughput of the \ac{MR} scheduler. Also note that at $\alpha=0$ the {LL-PF-Exp} is equivalent to a \ac{PF} scheduler operating with long-term user average rates only, as the short-term exponential utility is canceled out in \eref{eq:PF-LT}. We refer to this as the '\ac{PF}-Long-term' scheduler. The '\ac{PF}-Short-term' refers to the traditional \ac{PF} scheduler with a $W=1$\,s. Looking at \fref{fig:SlotvsAlpha}, we observe that with $\alpha=0$, the user short-term starvation metric $T^{\tn{10\%}}_{\tn{Slot}}$ is higher for the PF-Short-term as expected. As $\alpha$ increases, the figures jointly illustrate the throughput reduction and $T^{\tn{10\%}}_{\tn{Slot}}$ increase of the LL-PF-Exp scheduler. By appropriately selecting $\alpha$, the desired trade-off can be obtained. In addition to providing this trade-off the \ac{LLS} framework provides long-term fairness over \textit{multiple} cells as shown in \fref{fig:JvsAlpha}. This is illustrated in the long-term fairness measure of LL-PF-Exp, which is higher than the traditional PF-Short-term scheduler for all values of $\alpha$. This is due to the exchange of historical user rates between BSs, and its inclusion in the overall scheduling metric.

\begin{figure}[!t]
	\centering
	\includegraphics[trim=0cm 1.5cm 0cm 1.2cm, clip=true,width=75mm]{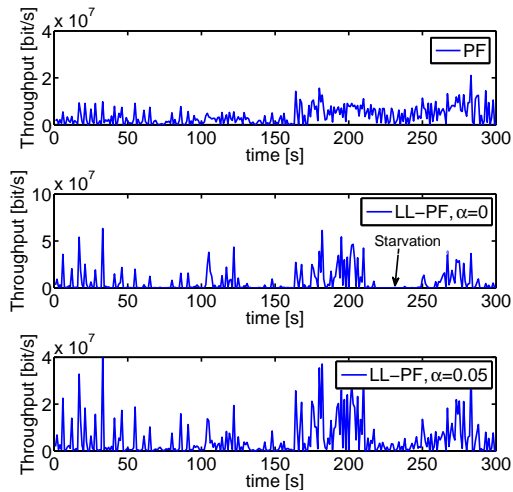}
	\caption{Sample user throughput vs time for \ac{PF} (Short-term) and \ac{LL-PF} schedulers.} 
	 %User speed is $S=40$\,km/h, and $N=250$}.
	\label{fig:TP_time}
\end{figure}
\fref{fig:Alpha_PF-LT} and \fref{fig:JvsAlpha} also compare the performance of LL-PF-Exp with various parameterizations of the LL-PF-Sig scheduler. As expected, LL-PF-Sig offers a broader trade-off with a particularly significant effect on $T^{\tn{10\%}}_{\tn{Slot}}$. With a high slope of $c=20$ and a relatively large $\beta$, the scheduler achieves a high $T^{\tn{10\%}}_{\tn{Slot}}$ (i.e. a low short-term starvation measure). This is due to sudden prioritization of users as their short-term rates start to decrease and approach the value of $\beta$. \fref{fig:JvsAlpha} also shows that the various LL-PF-Sig schedulers also achieve a higher long-term fairness compared to the single-cell short-term PF scheduler.

In \fref{fig:TP_time} we demonstrate the importance of preventing short-term starvation by showing a sample user throughput over time for the different schedulers. The traditional \ac{PF} scheduler has a $W=1$\,s and achieves a minimal throughput per second for the user (which comes at the cost of a reduced network throughput). On the other hand, LL-PF-Exp with $\alpha=0$ operates like a PF-Long-term only scheduler with $W=300$\,s which results in times where the user is not served at all and is starved (leading to a low value of the proposed $T^{\tn{10\%}}_{\tn{Slot}}$ metric). When $\alpha=0.05$, LL-PF-Exp provides a balance between guaranteeing short-term data rates, and providing bursts of high data. The $\beta$ parameter of LL-PF-Sig has the same effect. The proposed \ac{LLS} framework achieves this trade-off, while simultaneously providing a higher long-term fairness measure.

\subsection{Performance of \ac{LL-EXP}}
When studying the performance of the \ac{LL-EXP} schedulers, we set $a_i=1$ to assure equal priority to all users. We also set $R_\textnormal{Stream}=1.5\,\textnormal{Mbps}$.
We simulate two cases for the traffic arrival rate at the \ac{BS} from the core-network: $\lambda_i=12\,\tn{Mbps}$ and $\lambda_i=20\,\tn{Mbps},\forall i$.
The user terminal buffer playback threshold is set to 5 seconds. These parameters were chosen to simulate users streaming stored videos. Our metrics of interest are the average network throughput $T_{\tn{Net}}$, the average long-term video freezing $F^\tn{LT}$ experienced by the users, and the fairness in video freezing.

\begin{figure*}
    \centering
	\subfigure[Average network throughput $T_{\textnormal{Net}}$.]{
	\includegraphics[width=75mm]{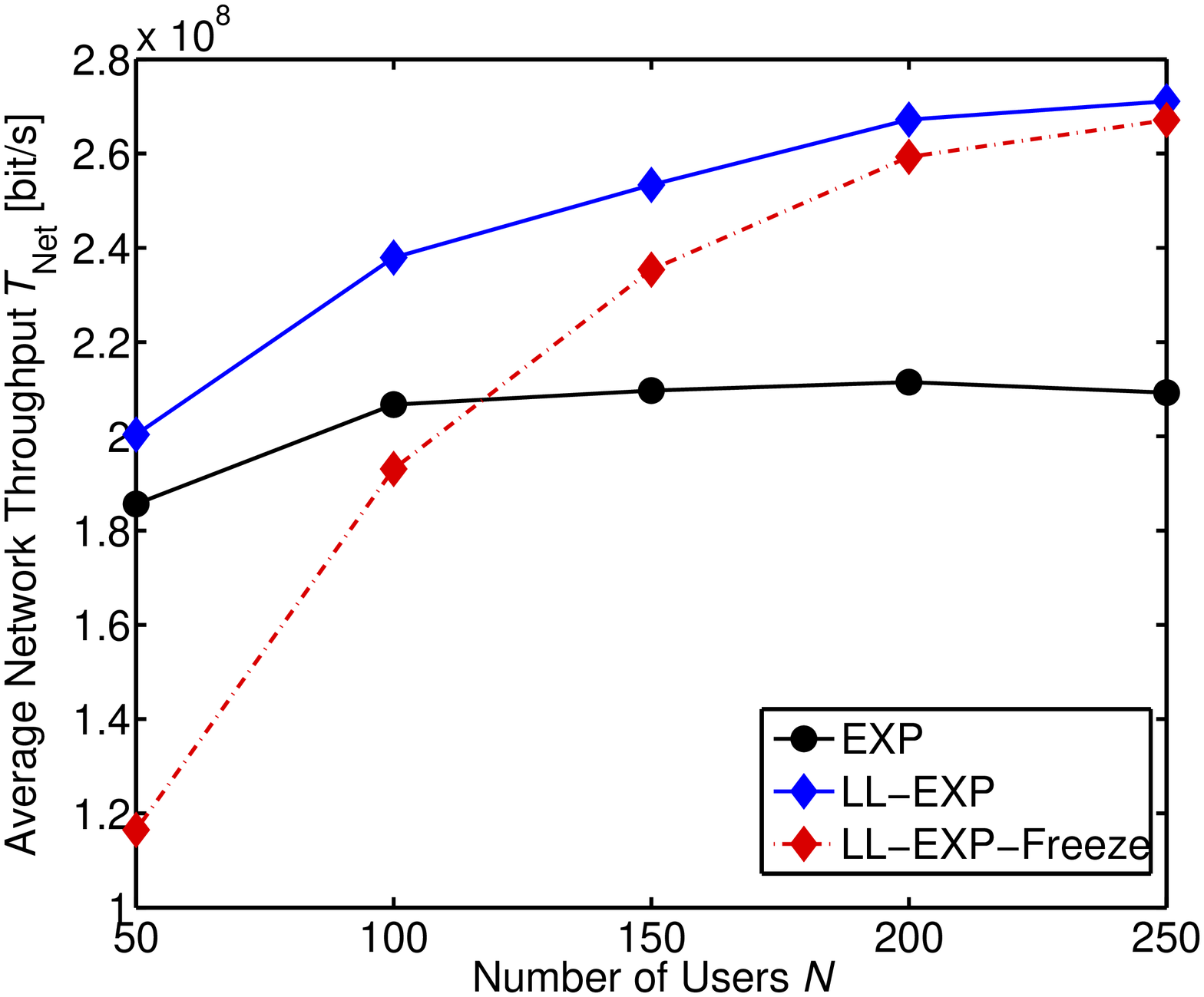}\label{fig:TvsN_Str_CN8}}
	\,
	\subfigure[Average network throughput $T_{\textnormal{Net}}$.]{
	\includegraphics[width=75mm]{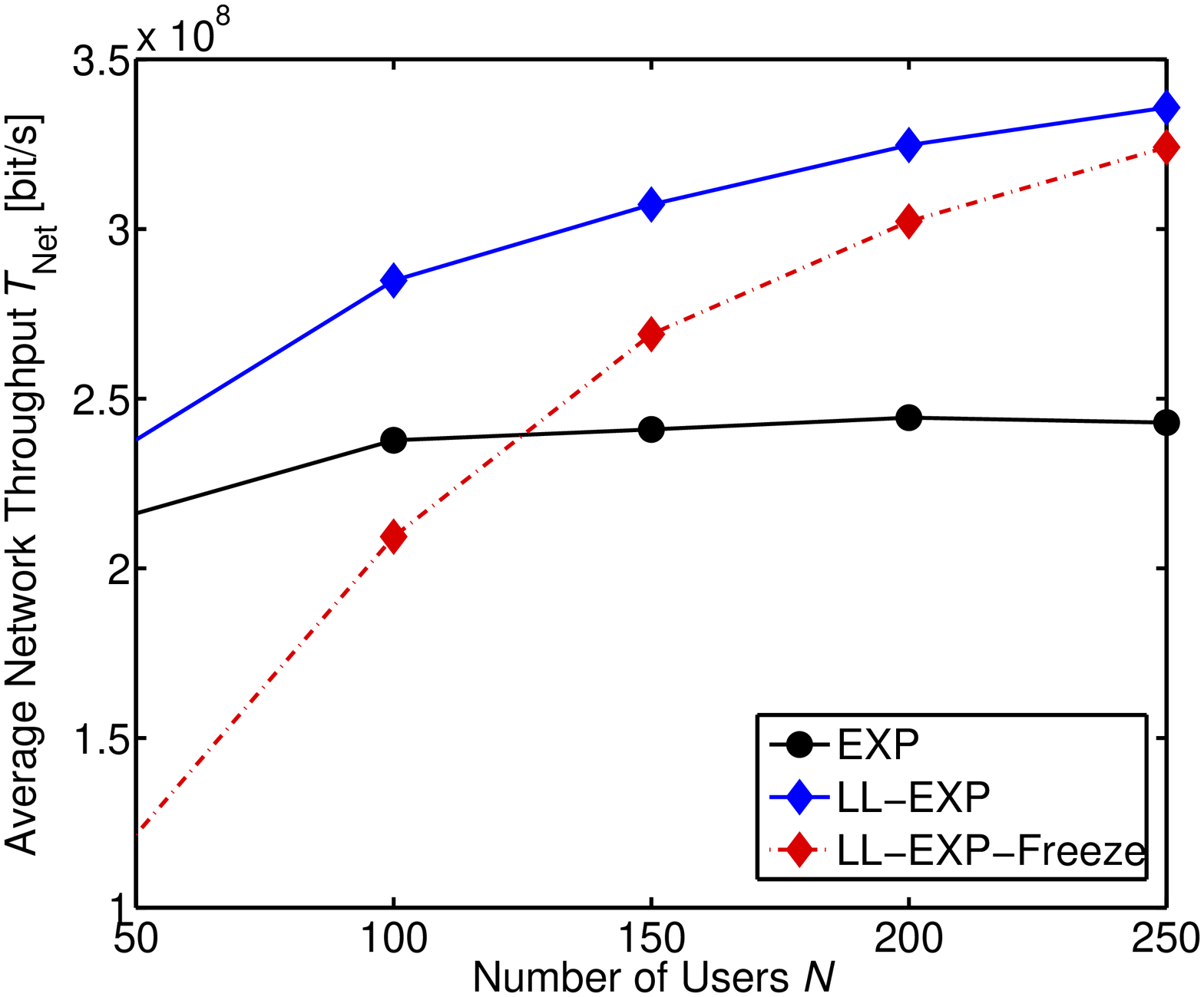}\label{fig:TvsN_Str_CN20}}
	\subfigure[Average \% of freezing $F^{\textnormal{LT}}$.]{
	\includegraphics[width=75mm]{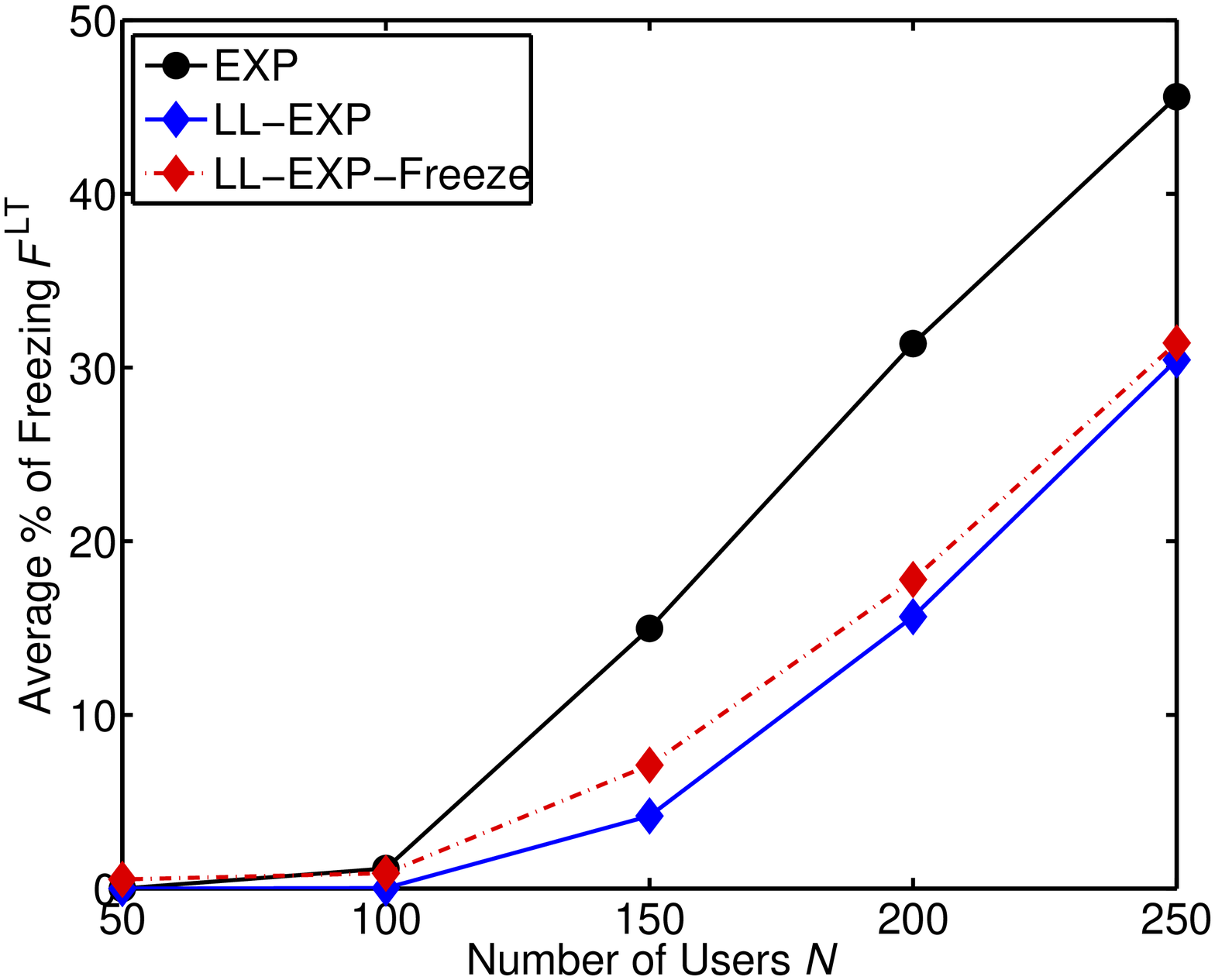}\label{fig:FvsN_Str_CN8}}
		\,
	\subfigure[Average \% of freezing $F^{\textnormal{LT}}$.]{
	\includegraphics[width=75mm]{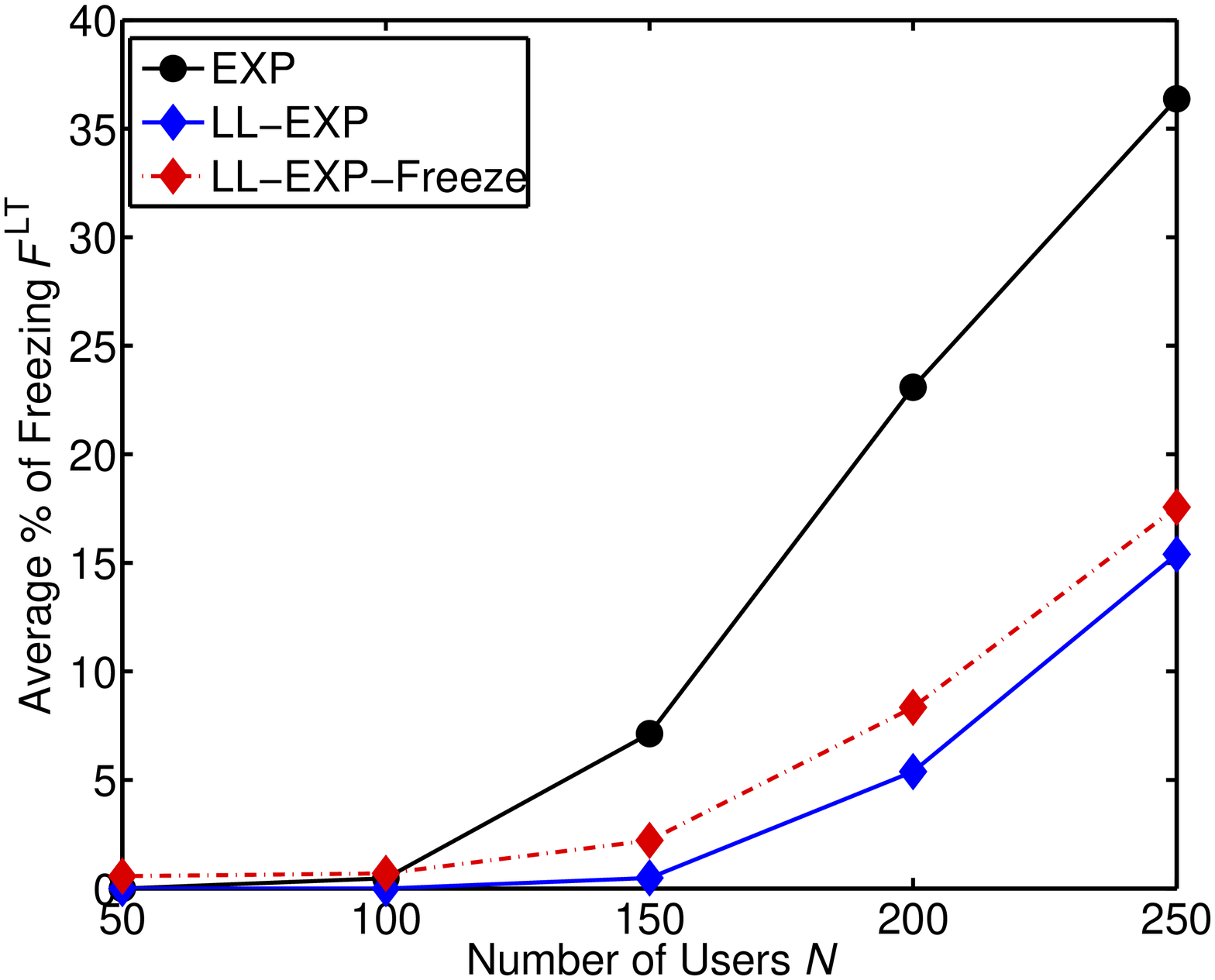}\label{fig:FvsN_Str_CN20}}
	\subfigure[Jain's fairness in freezing.]{
	\includegraphics[width=75mm]{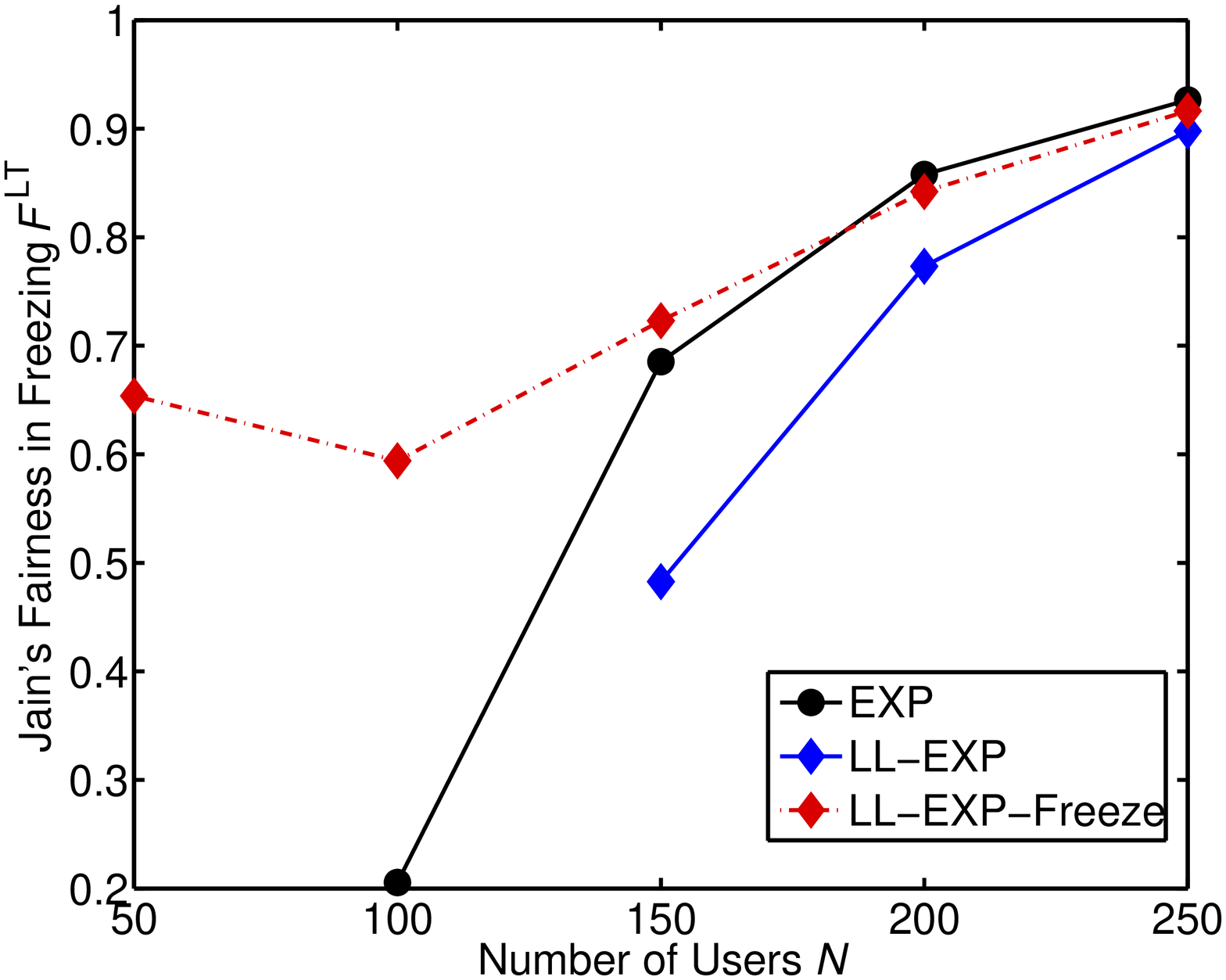}\label{fig:FJvsN_Str_CN8}}
	\,
	\subfigure[Jain's fairness in freezing.]{
	\includegraphics[width=75mm]{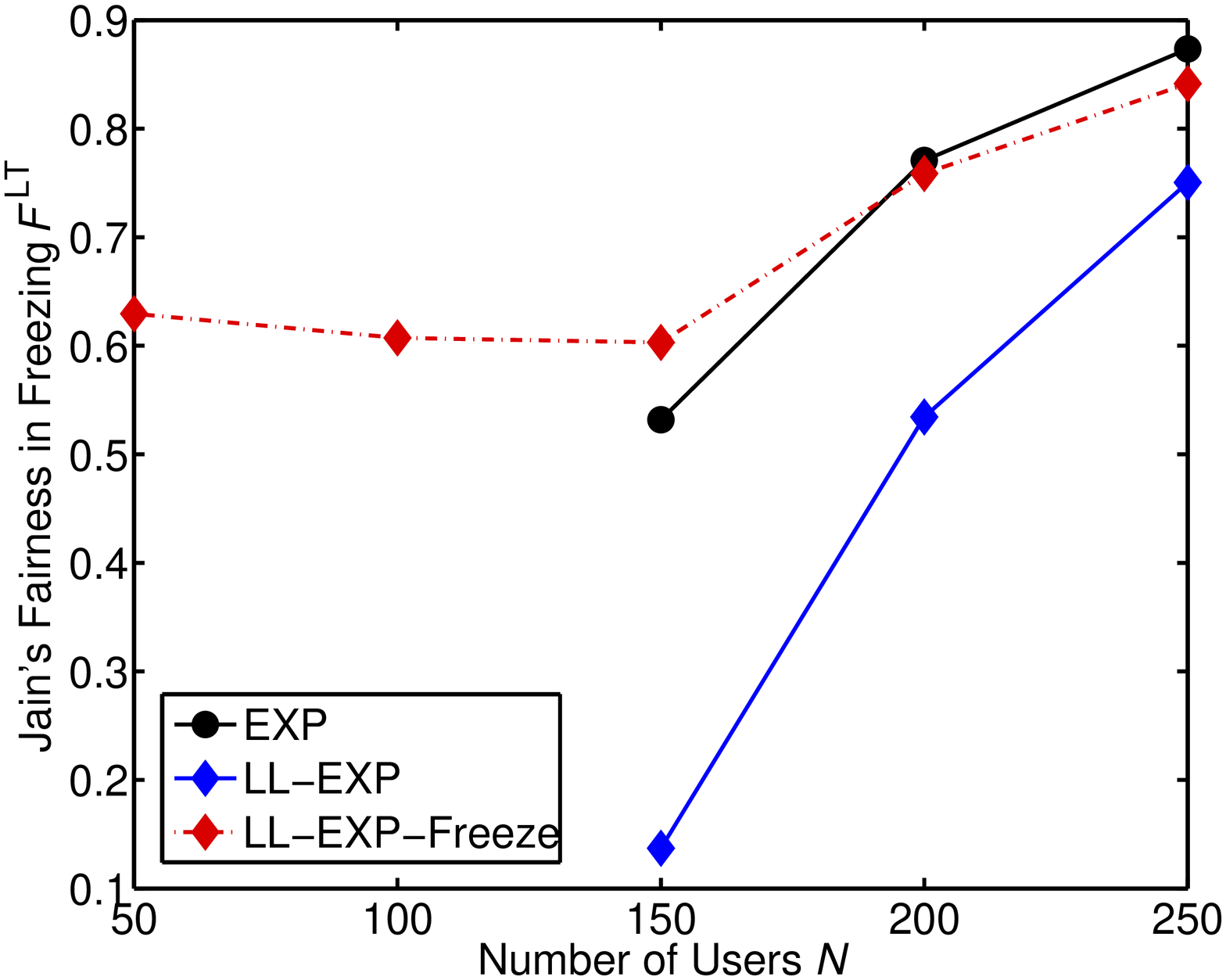}\label{fig:FJvsN_Str_CN20}}
	\caption{Effect of the number of users $N$ on the network throughput $T_{\textnormal{Net}}$, average video freezing $F^{\textnormal{LT}}$, and Jain's fairness in freezing. User speed $S=40$\,km/h, and $\lambda_i=12$\,Mbps in (a),(c),(e), and $\lambda_i=20$\,Mbps in (b),(d),(f).\\ Note that the discontinuities in (e) and (f) are for cases where there is no video freezing, and therefore there is no freezing fairness measure.}
	\label{fig:TFEXPvsN}
\end{figure*}

We first compare the LL-EXP to the EXP scheduler with a traffic arrival rate $\lambda_i=12\,\tn{Mbps}$. \fref{fig:TvsN_Str_CN8} shows the network throughput $T_{\tn{Net}}$ where we can observe a gain that increases with network load for \ac{LL-EXP} over the EXP scheduler. This gain arises as a consequence of the long-term average $R^\tn{LT}$ which allows the scheduler to opportunistically exploit good channel conditions of users even when other users have had a low short-term data rate. On the other hand, the traditional \ac{EXP} scheduler has two short-term indicators in its scheduling criterion as presented in \eref{eq:EXP}. This causes the scheduler to have excessive emphasis on achieving a short-term data rate for each user, and thereby prevents the scheduler from opportunistically serving users with high channel conditions. \fref{fig:TvsN_Str_CN8} illustrates this, where we observe that the \ac{EXP} scheduler throughput saturates quickly with an increasing number of users. In \fref{fig:FvsN_Str_CN8} we also see the significant reduction in freezing of the \ac{LL-EXP} scheduler, as it is able to support up to 160 users at a $F^\textnormal{LT}<5\%$ whereas \ac{EXP} can only support around 115. Therefore, the \ac{LL-EXP} scheduler can improve throughput without sacrificing user experience. 
Thus, the proposed \ac{LL-EXP} scheduler provides both throughput and long-term user experience gains. 

\fref{fig:FvsN_Str_CN8} shows how the LL-EXP-Freeze scheduler also reduces video freezing significantly, albeit at a lower rate than the LL-EXP scheduler. This is due to its explicit emphasis in providing fairness in video freezing as illustrated in \fref{fig:FJvsN_Str_CN8} which is a consequence of including the freezing amount in the scheduling metric of \eref{eq:EXP-F}). This fairness is also at the cost of a reduced throughput, as shown in \fref{fig:TvsN_Str_CN8}. 

In \fref{fig:TvsN_Str_CN20}, \fref{fig:FvsN_Str_CN20}, \fref{fig:FJvsN_Str_CN20} we present the throughput, freezing, and freezing fairness metrics respectively, for a similar setting but here traffic arrival rate $\lambda_i=20\,\tn{Mbps}$. In this case, the reduction in video freezing is even greater for both LL-EXP schedulers as shown in \fref{fig:FvsN_Str_CN20}. This is due to the constant availability of video content at the BS, which allows pre-buffering when the user channel conditions are good. The constant freezing fairness performance of the LL-EXP-Freeze scheduler is also observed in \fref{fig:FJvsN_Str_CN20}, with an even higher fairness gain compared to LL-EXP in this case.
%The EXP scheduler is not able to exploit this prebuffering due to it's emphasis on the 

\section{Conclusion}
\label{sec:Conc}
This paper introduced a \acf{LLS} framework to improve the \ac{QoS} of users traversing multiple cells. Our framework extends traditional schedulers by computing weights from information that was acquired over large time windows and previously visited cells. The choice of various utility functions (and their effects) for the short and long-term \ac{QoS} indicators was discussed to provide a direction for future research in long-term scheduling. LLS also requires no central coordination and adds only insignificant signaling overhead. 
The introduced notion of long-term service can also be applied to users that remain within a single-cell, to monitor and achieve long-term \ac{QoS}. Such a scheme can help avoid user frustration and improve subscriber retention in times of high traffic demand.

Our simulation results show that \ac{LLS} can significantly improve user satisfaction for both channel-aware and joint channel-queue-aware schedulers. We expect long-term scheduling to be a key approach for enabling constant high user satisfaction with small cells and uneven traffic distribution.

\ack This work was made possible by a \emph{National Priorities Research Program} (NPRP) grant from the Qatar National Research Fund (Member of Qatar Foundation).

\bibliography{refLLS}

\biogs
\textbf{Hatem Abou-zeid} 
is a postdoctoral fellow at the School of Computing, Queen's University. He received the Ph.D. degree in electrical and computer engineering from Queen's University, Canada, in 2014. During his program, he was awarded a DAAD RISE-Pro Research Scholarship in 2011 for a six-month internship at Bell Labs, Germany. 
Hatem is also an experienced Lecturer and has been granted several Teaching Fellowships at Queen's University to instruct freshman and senior-level engineering courses. His research interests include context-aware radio access networks, network adaptation and cross-layer optimization, adaptive video delivery, and vehicular communications. He is also a Technical Reviewer for several prestigious conferences and journals.
\bigbreak
\noindent\textbf{Hossam S. Hassanein} is the Founder and Director of the Telecommunications Research Laboratory, School of Computing, Queen's University, Canada, with extensive international academic and industrial collaborations. He is a leading
authority in the areas of broadband, wireless and mobile
networks architecture, protocols, control and performance
evaluation. His record spans more than 500 publications in
journals, conferences and book chapters, in addition to
numerous keynotes and plenary talks in flagship venues.
Dr. Hassanein served as the Chair for the IEEE Communication Society Technical Committee on Ad hoc and Sensor Networks (TC AHSN). He is a senior member of the IEEE, and an IEEE Communications Society Distinguished Speaker
(Distinguished Lecturer 2008-2010).
He has received several recognitions and Best Paper awards at top international conferences. 

\bigbreak
\noindent\textbf{Stefan Valentin} 
has been a full researcher at Bell Labs, Stuttgart, Germany since 2010. Previous appointments include the University of Paderborn, Germany, the International Centre of Theoretical Physics, Italy, and the Carleton University, Canada. Stefan's main research interest is context-aware wireless resource allocation with anticipatory adaptation and resource sharing in particular. In these fields, he leads several research projects and an integration activity with a major operator. Stefan received a summa cum laude doctorate in Computer Science from the University of Paderborn in 2010, the SIMUTools Best Paper Award in 2008, the Klaus Tschira Award for Comprehensible Science in 2011, and the Bell Labs Special Award of Excellence in 2013. He is a member of the Alcatel-Lucent Leadership program, advises the German Federal Ministry of Education and Research (BMBF), and leads the PhD Internship program at Bell Labs Germany.

\bigbreak
\noindent\textbf{Mohamed Feteiha} is an assistant professor and a researcher at the Informatic Research Institute (IRI) at the City of Scientific Research and Technological Applications (CSRTA), Alexandria, Egypt.  He is also a visiting researcher and previously a postdoctoral research fellow at the Telecommunications Research Lab (TRL) at Queen’s University, Ontario, Canada. He is working in the area of wireless networks architecture, deployments, and performance evaluation. Dr. Feteiha obtained his Ph.D. in
Electrical and Computer Engineering from University of Waterloo during 2008-2012, where he held a position of a research and teaching associate at the Wireless Communication Systems (WiComS) research lab. He received both his B.Eng. (Excellence grade with Honor) and M.Eng. (Excellence grade with Honor) degrees in Telecommunications Engineering from Faculty of Engineering, Arab Academy for Science and Technology (AAST), Alexandria - Egypt in 1999 and 2006, respectively. From 2000 to 2008, he was an assistant researcher at IRI-CSRTA. During this period, he intensively worked on distributed and wireless systems and networks.

\end{document}